\documentclass[aps,twocolumn,showpacs]{revtex4}
\usepackage{amsfonts}
\usepackage{amsmath}
\usepackage{amssymb}
\usepackage{graphicx}
\usepackage{multirow}

\def\bbbc{{\mathchoice {\setbox0=\hbox{$\displaystyle\rm C$}\hbox{\hbox
to0pt{\kern0.4\wd0\vrule height0.9\ht0\hss}\box0}}
{\setbox0=\hbox{$\textstyle\rm C$}\hbox{\hbox
to0pt{\kern0.4\wd0\vrule height0.9\ht0\hss}\box0}}
{\setbox0=\hbox{$\scriptstyle\rm C$}\hbox{\hbox
to0pt{\kern0.4\wd0\vrule height0.9\ht0\hss}\box0}}
{\setbox0=\hbox{$\scriptscriptstyle\rm C$}\hbox{\hbox
to0pt{\kern0.4\wd0\vrule height0.9\ht0\hss}\box0}}}}

\begin{document}

\title{Cooper pairs in atomic nuclei}

\author{G. G. Dussel$^{1}$, S. Pittel$^{2}$, J.~Dukelsky$^{3}$, and P. Sarriguren$^{3}$}
\address{
$^{1}$ Departamento de Fisica J. J. Giambiagi, Universidad de
Buenos Aires, 1428 Buenos Aires, Argentina \\
$^{2}$Bartol Research Institute and Department of Physics and
Astronomy, University of Delaware, Newark, DE 19716 USA \\
$^{3}$ Instituto de Estructura de la Materia, CSIC, Serrano 123,
28006 Madrid, Spain }
\date{Received \today }

\begin{abstract}
We consider the development of Cooper pairs in a self-consistent
Hartree Fock mean field for the even Sm isotopes. Results are
presented at the level of a BCS treatment, a number-projected BCS
treatment and an exact treatment using the Richardson ansatz.
While projected BCS captures much of the pairing correlation
energy that is absent from BCS, it still misses a sizable
correlation energy, typically of order $1~MeV$. Furthermore,
because it does not average over the properties of the fermion
pairs, the exact Richardson solution permits a more meaningful
definition of the Cooper wave function and of the fraction of
pairs that are collective.
\end{abstract}

\author{}
\maketitle

The first breakthrough in the derivation of a microscopic theory
of superconductivity was the demonstration by Cooper \cite{Cooper}
in 1956 that bound pairs could be produced in the vicinity of the
Fermi surface for an arbitrarily small attractive interaction.
This was followed soon thereafter by the development of the BCS
theory \cite{BCS}, in which superconductivity was described as the
condensation of a set of correlated pairs averaged over the whole
system. Soon after the BCS paper, Bohr, Mottelson and Pines
\cite{BMP} suggested that a similar phenomenon could explain the
large gaps in the spectra of even-even nuclei. Since then, the BCS
theory has been widely used to describe superconductivity in
condensed matter and nuclear systems. Moreover, the concept of
Cooper pairs as strongly overlapping objects that go through a
condensation process at the superconducting transition is central
in the interpretation of the superconducting phenomenon. However,
it is not easy to define the Cooper pair wave function from the
mean field BCS theory, and most frequently it has been related to
the pair correlator.

By using the exact solution of the BCS Hamiltonian given by Richardson in the sixties \cite{Richardson}, it was
recently shown \cite{Ortiz} that the Cooper pair wave function in  a superconducting medium  has a precise
definition. The unique form of its wave function transforms from a Cooper resonance in the weak coupling BCS
region to a quasi bound pair in the Bose-Einstein condensed (BEC) phase. Moreover, the Richardson solution gives a
clear prescription for evaluating the fraction of correlated pairs as compared with Yang's definition \cite{Yang},
 providing a more accurate description of the condensation phenomenon.

The subject of Cooper pairing in atomic nuclei has come under
renewed focus recently in the context of the mean-field
Hartree-Fock-Bogolyubov approach \cite{Matsuo, Pillet}. In this
work, we also explore the role of Cooper pairs in mean-field
treatments of atomic nuclei, comparing the traditional
number-nonconserving BCS approach with a projected BCS approach
and the exact Richardson treatment. We show that substantial
differences in correlation energies arise when pairing is treated
exactly for the same pairing strength, and that interesting
differences emerge in some conceptual properties of the paired
system.

We begin by detailing the differences between the three
approaches, focusing on a pairing Hamiltonian with constant
strength $G$ acting in a space of doubly-degenerate time-reversed
states $(k,\bar{k})$,
\begin{equation}
H=\sum_{k } \epsilon_k c^{\dagger}_{k } c_{k } -G \sum_{k,k'}
 c^{\dagger}_{k} c^{\dagger}_{\overline{k} }
c_{\overline{k}'} c_{k'} ~,\label{HBCS}
\end{equation}
where $\epsilon _k$ are the single-particle energies for the doubly-degenerate orbits $k,\bar{k}$.

Cooper studied the problem of adding a pair of fermions with an
attractive pairing interaction on top of an inert Fermi sea (FS).
He showed that the pair eigenstate is

\begin{equation}
\left\vert \Psi _{Cooper} \right\rangle = \sum_{k>k_{F}}\frac{1}{%
2\epsilon _{k}-E} ~c_{k}^{\dagger }c_{\overline{k}}^{\dagger
}\left\vert FS\right\rangle ~,\label{Cooper}
\end{equation}
where $E$ is the energy eigenvalue. It turns out that $E$ is negative for any attractive value of $G$, implying
that the Cooper pair is bound and that the FS is unstable against the formation of bound pairs. Cooper suggested
\cite{Cooper} that a theory considering a collection of bound pairs on top of an effective FS could explain
superconductivity.

The BCS approach follows a somewhat different path, defining
instead a variational wave function as a  coherent state of pairs
properly averaged over the whole system,

\begin{equation}
\left\vert \Psi _{BCS}\right\rangle = e^{\Gamma ^{\dagger }}\left\vert 0\right\rangle ~, \label{BCS}
\end{equation}
where $\Gamma ^{\dagger }=\sum_k z_k c^{\dagger}_{k}
c^{\dagger}_{\overline{k}}$ is the coherent pair. The BCS wave
function breaks particle-number conservation. Though errors due to
the nonconservation of particle number are negligible in the
thermodynamic limit,  they can be important in finite systems such
as atomic nuclei. Indeed, Bohr, Mottelson and Pines \cite{BMP}
noted already in 1958 the importance of taking into account finite
size effects in its application to nuclei. To accommodate these
effects, the number-projected BCS formalism (PBCS) \cite{DMP}
assumes a condensed state of pairs of the form

\begin{equation}
\left\vert \Psi _{PBCS}\right\rangle = \left( \Gamma ^{\dagger
}\right) ^{M}\left\vert 0\right\rangle ~,\label{PBCS}
\end{equation}
where $M$ is the number of pairs and  $\Gamma ^{\dagger }$ has the same form as in BCS. We would like to emphasize
here that $\Gamma ^{\dagger }$ should not be confused with the operator that creates a Cooper pair since its
structure contains an average over the correlated pairs close to the Fermi energy and the free fermions deep
inside the Fermi sphere.

The Richardson ansatz \cite{Richardson} for the exact solution of the pairing Hamiltonian (\ref{HBCS}) follows
closely Cooper's original idea. For a system with $2M$ particles, it involves (in the $\nu=0$  sector) a product
of $M$ distinct  pairs of the form

\begin{equation}
|\Phi>= \prod_{\alpha=1} ^M \Gamma^{\dagger}_{\alpha} |~0\rangle ~, ~~~\Gamma^{\dagger}_{\alpha} =  \sum_{k}
\frac{1}{2\epsilon_k-e_{\alpha}} ~ c^{\dagger}_{k} c^{\dagger}_{\overline{k}}~. \label{Ans}
\end{equation}

%where

%\begin{equation}
%\Gamma^{\dagger}_{\alpha} =  \sum_{k}
%\frac{1}{2\epsilon_k-e_{\alpha}} ~ c^{\dagger}_{k}
%c^{\dagger}_{\overline{k}}~.
% \label{Coppair}
%\end{equation}

The $e_{\alpha}$, called pair energies in analogy with the Cooper
wave function (\ref{Cooper}), are in general complex parameters,
which are obtained by solving the set of coupled non-linear
Richardson equations
\begin{equation}
1- G \sum_k \frac{1}{2\epsilon_k - e_{\alpha}}
 -2 G \sum_{\beta (\neq
\alpha) =1, M} \frac{1}{e_{\beta} -e_{\alpha}}=0 ~.\label{Req}
\end{equation}
The energy eigenvalues are obtained by summing the lowest $M$ pair
energies of each independent solution ($E=\sum_{\alpha}
e_{\alpha}$).

The key point to note upon inspection of the Richardson pair (\ref{Ans}) is that a pair energy close to a
particular $2 \epsilon _k$, i.e. close to the energy of an unperturbed pair, is dominated by this particular
configuration and thus defines an uncorrelated pair. In contrast, a pair energy lying sufficiently far away in the
complex plane produces a correlated Cooper pair.

As mentioned before, the BCS coherent pairs, with  amplitudes
$z_k=v_k/u_k$, cannot be interpreted as Cooper pairs since they
mix correlated and uncorrelated pairs over the whole system.
Indeed, it has been shown \cite{Ortiz} that only in the extreme
BEC limit are all pairs bound and condensed, and amenable to
description by the two approaches. Usually the structure of the
Cooper pair  is assigned to the pair correlator $\left\langle
BCS\right\vert c_{k}^{\dagger }c_{\overline{k}}^{\dagger
}\left\vert BCS\right\rangle=u_k v_k $. However, if the BCS state
represents a fraction of correlated pairs within a Fermi sea of
free uncorrelated fermions, the pair correlator cannot guarantee
that it picks up the two fermions from the same pair. The pair
correlator is another averaged property over the set of correlated
pairs.

In what follows we explore the structure of pairing correlations
in the even Sm isotopes, from $^{144}$Sm through $^{158}$Sm. The
results are based on a series of self-consistent deformed Hartree
Fock+BCS calculations. The calculations make use of the
density-dependent Skyrme force, SLy4, and treat pairing
correlations using a pairing force with constant strength $G$.

The calculations are carried out in an axially symmetric harmonic oscillator space of 11 major shells (286
doubly-degenerate single-particle states). This basis involves oscillator parameters $b_0$ and axis ratio $q$,
optimized in order to minimize the energy in the given space. The strength of the pairing force for protons and
neutrons is chosen in such a way as to reproduce the experimental pairing gaps in $^{154}$Sm ($\Delta_n=0.98$
$MeV$, $\Delta_p= 0.94$ $MeV$), extracted from the binding energies in neighboring nuclei. We obtain $G_n=0.106$
$MeV$ and $G_p=0.117$ $MeV$. Once we have fitted this reference strength, we determine the pairing strengths
 appropriate to the $^{142-158}$Sm isotopic chain  by
assuming a $1/A$ dependence. These calculations provide an
excellent description of the properties of the even Sm isotopes.

We then use the results at self-consistency to define the HF mean field and consider the alternative
number-conserving PBCS and exact Richardson approach to treat the pairing correlations within {\it this} mean
field. We ignore the issue of whether the mean field should be self-consistently modified in these other
approaches. In this way we are able to directly compare the three approaches to pairing with the same pairing
Hamiltonian, which is the focus of this investigation.

As is well known that the numerical solution of the Richardson equations (\ref{Req}) involves instabilities due to
singularities arising at some critical values of the pairing strength $G$. There have been two recent works that
study these critical regions of parameter space \cite{Fer} and propose ways to overcome the singularities
\cite{Rom}.  While these methods alleviate the numerical divergences, thus allowing for an interpolation method to
cross the critical regions, some problems still persist and we have thus chosen to use a different approach. Since
the singularities arise as crossings of real pair energies $e_{\alpha}$ with the unperturbed single-pair energies
$2\epsilon_k$ in the denominators of (\ref{Req}), we start the numerical procedure at strong coupling ($G=1~MeV$)
with complex single-particle energies, obtained by adding a small arbitrary imaginary component.  In this way, the
singularities are avoided in the evolution of the system from strong coupling almost to the $G=0$ limit.  To
obtain the exact solution at the physical value of $G$, we then let the imaginary parts go to zero starting with
the solution already obtained for that $G$ value. The method seems to work for any distribution of single-particle
energies.

A principal focus of our investigation is on the pairing correlation energy, defined as
\begin{equation}
E_C=\langle \Phi_{corr}|H|~\Phi_{corr}\rangle -\langle
\Phi_{uncorr}|H|~\Phi_{uncorr}\rangle~,
\end{equation}
where $|\Phi_{corr} \rangle$ is the correlated ground-state wave
function and $|\Phi_{uncorr} \rangle$ is the uncorrelated Hartree
Fock Slater determinant obtained by filling all levels up to the
Fermi energy.   This quantity reflects the additional energy that
derives from the inclusion of pairing.

Table 1 summarizes our results for the pairing correlation energy in table 1 for all the even Sm isotopes under
consideration. Note that the calculations include the semi-magic nucleus $^{144}$Sm, for which the BCS calculation
leads to a normal solution with no pairing correlation energy. In contrast, the projected BCS calculation leads to
substantial pairing correlations in the ground state. That number projection is critical in mean-field treatments
of semi-magic nuclei is well known from other calculations \cite{Stoitsov}. The exact treatment of pairing leads
to a further lowering of the energy of the ground state of the system, by $0.3$ $MeV$.

In the calculations other than $^{144}$Sm, the effect on the pairing correlation energy of the exact solution is
more pronounced. While PBCS gives a significant lowering of the energy of the system due to number projection, it
misses about $1$ $MeV$ of the full correlation energy of an exact treatment. Considering the extensive recent
efforts to carry out systematic microscopic calculations of nuclear masses using mean-field methods
\cite{Pearson}, we feel that this effect may be quite meaningful. It is not clear that a renormalization of the
strength of the pairing interaction can accommodate these important corrections.

The results obtained for the Sm isotopes are consistent with studies performed in ultrasmall superconducting
grains \cite{Grain1}. The quantum phase transition from a superconducting to a normal metal predicted by BCS and
PBCS completely disappears after fully including the pairing fluctuations by means of the exact solution of the
BCS model. Moreover, the PBCS wave function displays a strange behavior in the transitional region as compared
with the smooth behavior of the exact wave function \cite{Grain2}.

\begin{center}
{Table I: Pairing correlation energies associated with the BCS,
PBCS and exact Richardson treatments
of pairing for the even Sm isotopes. All energies are given in $MeV$} \\
\end{center}
\begin{center}
\begin{tabular}{l|l|l|l|l|l}
\hline
&Mass& $E_C(Exact)$ & $E_C(PBCS)$ & $E_C(BCS)$ \\
\hline
&142& -4.146   &    -3.096    &   -1.107 \\
&144& -2.960   &   -2.677    &    ~0. \\
&146 & -4.340   &    -3.140   &    -1.384\\
&148 & -4.221    &    -3.014   &    -1.075\\
& 150 & -3.761   &    -2.932   &   -0.386\\
& 152 & -3.922   &    -2.957   &  -0.637\\
& 154 & -3.678    &   -2.859    &  -0.390 \\
& 156 &  -3.716   &    -2.832   &  -0.515 \\
& 158 & -3.832    &   -2.824    &  -0.717\\
 \hline
\end{tabular}
\end{center}

A second important feature of Cooper pairing is the condensate fraction, namely the fraction of pairs of  the
whole system that are correlated. Analysis of the off-diagonal long-range order (ODLRO) that characterizes
superconductors and superfluids led Yang \cite{Yang}  to a definition of the condensate fraction, $\lambda$, in
terms of the single macroscopic eigenvalue of the two-body density matrix. For a homogeneous system of two spin
fermion species in the thermodynamic limit, $\lambda$ is given by

\begin{equation}
\lambda =\int d^{3}\mathbf{r}_{1}d^{3}\mathbf{r}_{2}\left\vert
\left\langle \psi _{\downarrow }\left( \mathbf{r}_{1}\right) \psi
_{\uparrow }\left( \mathbf{r}_{2}\right) \right\rangle \right\vert
=\frac{1}{M}\sum_{k}u_{k}^{2}v_{k}^{2} ~.\label{Yang}
\end{equation}

This definition is not appropriate for finite Fermi systems,
however, where several eigenvalues of the two-body density matrix
are of the same order. We modify it, therefore, by excluding from
the two-body density matrix the amplitude of finding two
uncorrelated fermions. More specifically, our prescription for
finite systems is to evaluate the matrix elements of the operator
\begin{equation}
\lambda= \frac{1}{M(1-M/L)}  \sum_{k,k'=1}^{L} \langle
c^{\dagger}_{k} c^{\dagger}_{\bar{k'}} c_{\bar{k'}} c_{k} \rangle
- \langle c^{\dagger}_{k}c_{k}\rangle \langle
c^{\dagger}_{\bar{k'}}c_{\bar{k'}}\rangle ~,
\label{Yang_revised}
\end{equation}
where $L$ is the total number of doubly-degenerate, canonically
conjugate pair states $k,\bar{k}$.

%This prescription has a further difference from Yang's definition through the introduction of an extra term in the
%normalization to reflect the fact that we are treating a finite system with a cutoff.

In BCS approximation, the modified Yang prescription leads to a
condensate fraction

\begin{equation}
\lambda_{BCS} =\frac{1}{M(1-M/L)}\sum_{k=1}^L
u_{k}^{2}v_{\bar{k}}^{2} ~.
\end{equation}
We have calculated this quantity for the BCS solutions obtained for $^{154}$Sm  as a function of the pairing
strength $G$ and plot the results as the smooth curve in figure 1.

An alternative prescription for the condensate fraction from the exact Richardson solution was proposed in
\cite{Ortiz} and shown to more properly reflect the properties of a superfluid system as it undergoes the
crossover from BCS to BEC. In particular, this new prescription gives a fully condensed state at the change of
sign of the chemical potential where the whole system becomes bound. This prescription, however, requires
knowledge of the properties of the precise Cooper pairs in the problem, not an average over the whole system as
provided by the BCS or PBCS approximations (\ref{BCS},\ref{PBCS}). The Richardson ansatz (\ref{Ans}) is ideally
suited for this as it provides an exact wave function for each individual Cooper pair. One has to simply
distinguish which pairs are correlated and which are not. As previously discussed, a correlated pair is
characterized by a pair energy $e_{\alpha}$ that is far enough away in the complex plane from any particular $2
\epsilon _k$. We therefore propose the following practical definition for the condensate fraction. {\it It is the
fraction of pair energies which in the complex energy plane lie further from any unperturbed single-pair energy,
$2\epsilon_k$, than the mean single-particle level spacing.}

\begin{figure}[htb]
\includegraphics[height=.28\textheight]{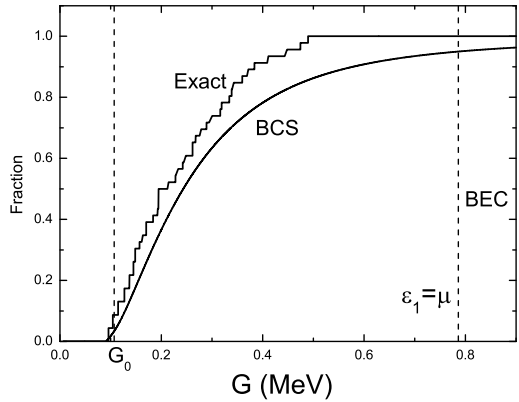}
\caption{ The modified Yang prescription for the BCS treatment of pairing (smooth curve) and the alternative
prescription discussed in the text (sawtooth curve) for the exact Richardson treatment. $G_0=0.106~MeV$ denotes
the physical value of the pairing strength and $\epsilon_1=\mu$ denotes the strength at which the whole system
binds.}
\end{figure}

We now return to a discussion of the condensate fraction, as
plotted in figure 1 for $^{154}$Sm as a function of the pairing
strength $G$.  In addition to the results based on the pair
correlator, as discussed earlier, we also plot (in the sawtooth
curve) the results that derive from the exact Richardson solution
using the prescription just described. To illustrate how these
latter results emerge, we show in figure 2 the associated pair
energies for four values of $G$ in $^{154}$Sm, ranging from the
physical value of $G=0.106$ $MeV$ to a fairly strong pairing
strength of $G=0.4$ $MeV$. In $^{154}$Sm the mean level spacing
between the Hartree Fock single-particle levels is roughly
$0.5~MeV$, both around the Fermi surface and far from it. For
$G=0.106$ $MeV$, most of the pair energies lie very near the real
axis and quite close to at least one unperturbed single-pair
energy, $2\epsilon_k$. Two of them (which form a complex conjugate
pair) extend about $1~MeV$ in the complex plane, while another two
are marginally collective, lying roughly 0.5 $MeV$ from the
closest $2\epsilon_k$. The two most collective pairs, denoted
C$_1$ in the figure, each have a real energy of $-15.55~MeV$,
which is roughly twice the energy of the single-particle levels
just below the Fermi surface. This suggests that the first pairs
that become collective are indeed those built out of the valence
orbits. As $G$ increases, we see a gradual increase in the number
of collective pairs, which form an arc in the complex plane. As
can be seen from figure 1, by a pairing strength of roughly
$0.5~MeV$ all of the pairs of the system are correlated giving a
condensate fraction of $1$, even though the BEC regime has not yet
been reached. The BEC limit is realized when the chemical
potential $\mu$ crosses the lowest single-particle energy
$\epsilon _1$ at $G=0.788$ for $^{154}$Sm. At this point all pairs
are bound. However, the revised Yang prescription
(\ref{Yang_revised}) fails to predict a complete condensate at
this point, in the same way as it fails to do so in the
homogeneous case \cite{Ortiz}.

\begin{figure}[htb]
\includegraphics[height=.28\textheight]{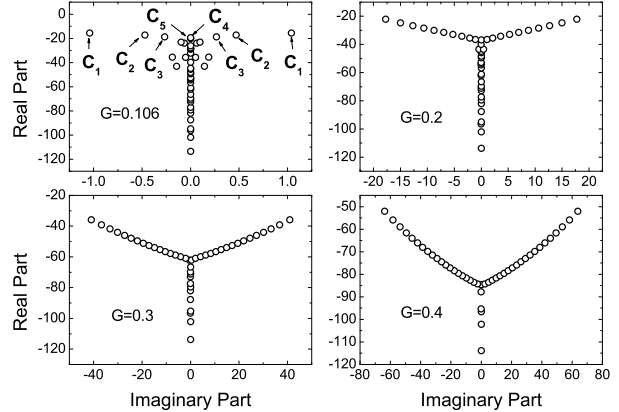}
\caption{Pair energies (in $MeV$) for the exact Cooper pairs that
emerge from four calculations of the $^{154}$Sm isotope. $G=$
$0.106~MeV$ is the physical value of the pairing strength. In that
panel, we denote the most collective pairs as C$_i$, for
subsequent notational purposes. }
\end{figure}

\begin{figure}[htb]
\includegraphics[height=.28\textheight]{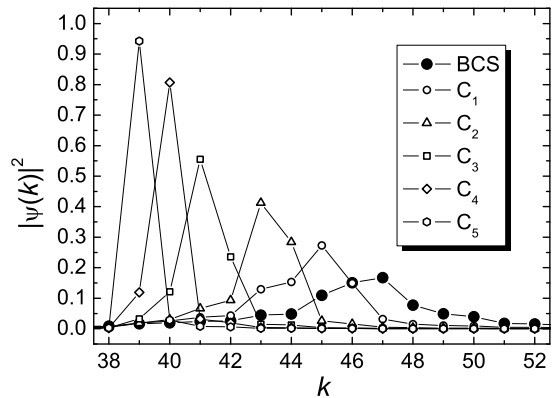}
\caption{Square of the wave function of the most collective Cooper pairs in $^{154}$Sm (denoted C$_1$, C$_2$,
C$_3$, C$_4$, and C$_5$) and the pair correlator (BCS) versus the single-particle levels.}
\end{figure}

The Richardson prescription for Cooper pairs also gives rise to a
different interpretation of their internal structure. In figure 3,
we compare the square of the wave function for the most correlated
Cooper pairs in $^{154}$Sm, {\it i.e.} those whose pair energies
lie farthest from any unperturbed single-pair energy, with the
square of the pair correlator wave function obtained from the BCS
calculation. All wave functions are plotted versus the order of
the single-particle states to make clear the relevant mixing of
configurations in each pair. The pair labels in the figure (C$_1$
through C$_5$) refer to corresponding labels in the upper left
panel of figure 2. C$_1$ refers to the two most collective pairs,
namely those that are farthest from any unperturbed
single-particle pair. Being complex conjugate pairs, both have
exactly the same absolute square of their wave function and thus
we only show one in the figure. C$_2$ refers to the next two most
collective pairs, which as noted earlier are marginally collective
according to our prescription. C$_3$ refers to the next two most
collective pairs after C$_2$, which according to the prescription
given above involve perturbative mixing of configurations and are
not collective. C$_4$ and C$_5$, the following pairs in descending
collective order, have real pair energies and involve almost pure
single-particle configurations.

>From the figure, we see that the pair correlator wave function is quite spread over several single-particle
configurations and is peaked between at the $47^{th}$ single-particle level, just beyond the Fermi energy
($^{154}$Sm has $46$ neutron pairs). In contrast, the most highly correlated Cooper pair wave function C$_1$ is
somewhat narrower (less collective) and is peaked slightly within the Fermi sphere. The less-collective Cooper
pairs, C$_2$ through C$_3$, are peaked progressively further inside the Fermi sphere and are progressively
narrower. From this figure, we conclude that the size of even the most collective Cooper pairs in coordinate space
will be larger than the size of the pair correlator, as was already demonstrated in the weak coupling BCS regime
of cold atomic gases \cite{Ortiz}. Recent investigations \cite{Matsuo, Pillet} on the size of the pair correlator
in spherical nuclei have concluded that it is unexpectedly small in the nuclear surface ($2-3 ~ fm$). The present
calculations would suggest that the actual size of the few highly collective Cooper pairs is larger than the
typical size of the pair correlations in the nuclear medium. Furthermore, as is also evident from the figure, less
bound pairs get progressively closer to a particular $2\epsilon_k$ and the corresponding Cooper pair wave function
is less collective, i.e. more narrow in energy space, and peaked at this particular configuration.

In this work, we have studied the role of Cooper pairing in atomic nuclei, focusing on a realistic description of
the even Sm isotopes. We assume that the mean field is given by the self-consistent HF solution from coupled
HF+BCS calculations, and then consider how the effects of pairing on that mean field would be modified at several
levels of improved treatment. We consider both the projected BCS approximation and an exact treatment based on
Richardson's solution of the pairing problem. Several important points emerged. On the one hand, even though PBCS
approximation gives a significant gain in binding energy over ordinary BCS, it still fails to capture a sizable
component, typically of order $1$ $MeV$. This might have important implications in efforts to derive nuclear
masses from a microscopic approach. Second, we discussed a new and improved prescription for identifying the
fraction of the pairs in a nucleus that are collective, which can only be realized when the properties of the
various Cooper pairs in the problem are treated separately. This new prescription suggests that a slightly larger
number of pairs are collective when compared to the more usual prescription based on Yang's definition of the
condensate fraction. Furthermore, it suggests that the few collective Cooper pairs that arise in real nuclei,
being individually less collective than  the pair correlator,  would be spatially more spread out.

The Richardson solution, as generalized in ref. \cite{DES}, can be obtained for integrable pairing hamiltonians
only. It is possible, however, to use the Richardson ansatz (\ref{Ans}) in a variational treatment of general
non-integrable pairing hamiltonians.  The pair energies would play the role of variational parameters within a
generalized Pfaffian pairing wave function \cite{nonintegrable}, making it possible to treat pair correlations in
a more precise manner for realistic nuclear systems.

We acknowledge fruitful discussions with N. Sandulescu, P. Schuck and W. Nazarewicz. This work was supported in
part by the Spanish DGI under grants FIS2005-00640 and FIS2006-12783-C03-01, in part by the US National Science
Foundation under grant \# 0553127, and in part by UBACYT X-053, Carrera del Investigador Cient\'ifico and PIP-5287
(CONICET-Argentina). One of the authors (SP) wishes to acknowledge the generous support and hospitality of the
CSIC in Madrid where much of his contribution to the work was carried out


\begin{thebibliography}{99}

\bibitem{Cooper} Leon N. Cooper, Phys. Rev. {\bf 104}, 1189 (1956).

\bibitem{BCS} J. Bardeen, L. N. Cooper and J. R. Schrieffer, Phys.
Rev. {\bf 108}, 1175 (1957).

\bibitem{BMP} A. Bohr, B. R. Mottelson and D. Pines, Phys. Rev. {\bf
110}, 936 (1958).

\bibitem{Richardson} R. W. Richardson, Phys. Rev. Lett. {\bf 3},
277 (1963); Phys. Rev. {\bf 141}, 949 (1966).

\bibitem{Ortiz} G. Ortiz and J. Dukelsky, Phys. Rev. A {\bf 72}, 043611 (2005).

\bibitem{Yang} C. N. Yang, Rev. Mod. Phys. {\bf 34}, 694 (1962).

\bibitem{Matsuo} Masayuki Matsuo, Kazuhito Mizuyama and Yasuyoshi
Seizawa, Phys. Rev. C {\bf 71}, 064326 (2005).

\bibitem{Pillet} N. Pillet, N. Sandulescu and P. Schuck, LANL
preprint \# nucl-th/0701086.


\bibitem{DMP} K. Dietrich, H. J. Mang and J. H. Pradal, Phys. Rev. {\bf 135}, B22 (1964).

%\bibitem{Skyrme} A. Chabanat et al. Nucl. Phys. A635 (1998) 231 [2] D.
%Vautherin, Phys. Rev. C7 (1973) 296.

\bibitem{Fer} F. Dominguez, C. Esebbag and J. Dukelsky, J. Phys. A: Math. Gen. {\bf 39}, 11349 (2006).

\bibitem{Rom} S. Rombouts, D. Van Neck and J. Dukelsky, Phys. Rev. C {\bf 69}, 061303 (2004).

\bibitem{Stoitsov}M.V. Stoitsov, J. Dobaczewski, R. Kirchner, W. Nazarewicz and J. Terasaki, LANL preprint \#
nucl-th/0610061.

\bibitem{Pearson} D. Lunney, J.M. Pearson, C. Thibault
Rev. Mod. Phys. {\bf 75}, 1021 (2003).


\bibitem{Grain1} J. Dukelsky and G. Sierra, Phys. Rev. Lett. {\bf 83}, 172 (1999).

\bibitem{Grain2} J. Dukelsky and G. Sierra, Phys. Rev. B {\bf 61}, 12302 (2000).

%\bibitem{DEP} J. Dukelsky, C. Esebbag and S. Pittel, Phys. Rev. Lett. {\bf 88}, 062501 (2002).

\bibitem{DES} J. Dukelsky, C. Esebbag and P. Schuck, Phys. Rev. Lett. {\bf 87}, 66403 (2001).

\bibitem{nonintegrable}  M. Bajdich, L. Mitas, G. Drobný, L. K. Wagner, and K. E. Schmidt, Phys. Rev. Lett. {\bf 96}, 130201
(2006).


\end{thebibliography}
\end{document}